# Nanomechanical inhomogeneities in CVA-deposited titanium nitride thin films: Nanoindentation and Finite Element Method Investigations

**Neeraj Kumar Sharma, Anchal Rana, O. S. Panwar, Abhimanyu Singh Rana, ***

Centre for Advanced Materials and Devices, School of Engineering and Technology, BML Munjal University, Sidhrawali, Gurugram-122413. Haryana, India

*rana.abhimanyu@gmail.com

**Abstract**

Refractory metals that can withstand at high temperatures and harsh conditions are of utmost importance for solar-thermal and energy storage applications. Thin films of TiN have been deposited using cathodic vacuum arc deposition (CVA) at relatively low temperatures ~ 300 $^o$C using the substrate bias ~ -60V. The nanomechanical properties of these films were investigated using nanoindentation and the spatial fluctuations were observed. The nanoindentation results were simulated using finite element method (FEM) through Johnson-Cook model. We have found the local nitridation plays an important role on nanomechanical properties of TiN thin films and confirms that the nitrogen deficient regions are ductile with low yield stress and hardening modulus. This study further opens the opportunities of modelling the nanoscale system using FEM analysis.

**Introduction**

There are not many refractory materials with both metallic and plasmonic characters that are suitable for thermo-photovoltaics and solar-thermal applications and that can replace the precious metals like silver and gold.[1–7] There is a renewed interest in titanium nitride (TiN) as a plasmonic metamaterial[4,8–11] as a cathode material for energy storage[12–15] and catalysis.[16–18] The nanomechanical properties, surface defects and stoichiometry are crucial for the overall device functionality and reliability [5,6,19–23]. Vacuum-based physical vapor deposition (PVD) techniques such as sputtering,[24,25] high-power impulse magnetron sputtering,[20] pulsed laser deposition,[26,27] and cathodic vacuum arc (CVA) deposition[28] have been widely used for synthesizing the TiN thin films. However, measuring the local properties of metal-nitrides is a daunting task as the inert nature of the nitrogen molecules and the undesirable formation of oxides can lead to nanoscopic inhomogeneities in mechanical properties. The nanoindentation technique can be a promising method to investigate the nanomechanical properties.[29] During an indentation test, the applied force and the indentation depth is continuous monitored and the elastic modulus and the hardness can be measured. In this work, the titanium nitride (TiN) thin films grown by CVA deposition have been investigated using the nanoindentation to understand the nanomechanical properties of titanium nitride (TiN). Furthermore, the finite element method (FEM) has also been implemented simulate the nanoindentation results to understand the elasto-plastic behaviour and predict the yield stress, strain hardening coefficient and the strain hardening exponent.

**Experimental**

The CVA deposition of TiN thin films were carried out using a titanium target (75 mm diameter and ~ 99.999% purity) on well cleaned 7059 glass and silicon substrates at an arc current of ~160A, keeping a negative substrate bias of -60V, at a temperature of ~ 300⁰C under different nitrogen background pressures ~ 1 x$10^{-3}$ mbar without the use of any magnetic filter. Prior to the deposition, a base pressure better than 5 x $10^{-6}$ mbar was achieved, and the substrates were cleaned *in-situ* using the argon gas plasma for 5 minutes at the DC bias voltage ~ - 460V. The crystal structure of the films was confirmed by X-ray diffraction (XRD) by Panalytical and Raman spectroscopy by Witec (Alpha 300) system. Raman spectra were acquired using a green laser (532 nm) with a ~ 5 mW incident power. The chemical composition and stoichiometry of these films were confirmed by X-ray photoelectron spectroscopy (XPS) using Thermo Fisher (k Alpha) system. The load – displacement curves of the films have been measured using IBIS nanoindentation (Fischer-Cripps Laboratories Pvt. Ltd., Australia) having triangular pyramid diamond Berkovich indenter with a normal angle of 65.3° and tip radius of ~100 nm. The surface was analysed by the scanning electron microscope by Hitachi (SU3500). Thickness and the roughness of these thin films were measured by surface profilometer by Brucker, Dektak XT.

**Results and discussion**

**Figure 1 (a)** shows the XRD pattern of TiN films deposited using the cathodic arc deposition technique. The XRD pattern shows sharp and intense peaks at ~36.2º, 42.3º, and 61.27º, which correspond to the (111), (200), and (220) planes, respectively, indicating that TiN has a cubic crystal structure. The peaks at 74º and 77.8º correspond to the (311), and (222) planes respectively. In addition to that, these peak are slightly shifted to a higher angles due to compressive-strain that could be originated due to the lattice mismatch between the Ti-rich caused by nitrogen vacancies.[30]

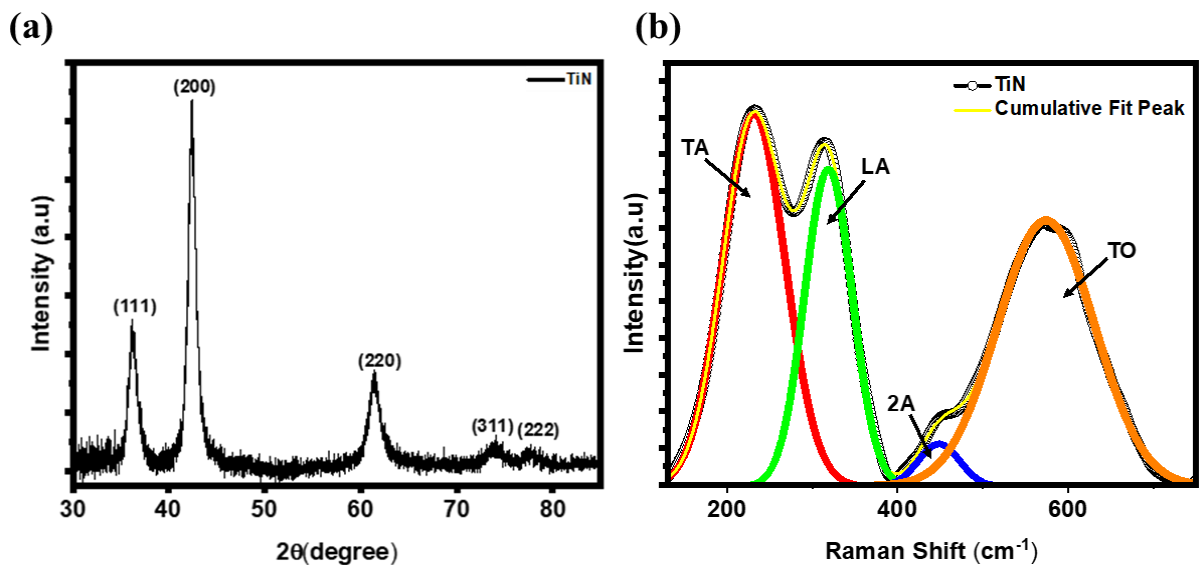

**Figure 1: (a) X-ray diffraction (XRD) and (b) Raman spectroscopy results of CVA-deposited TiN thin films.**

Also, TiN is one of the transition metal nitrides that does not show first-order Raman scattering due to $O_h$ symmetry. But the presence of defects (or vacancies) within the lattice induces Raman scattering. We also observed the Raman peaks in these films as shown in figure 1 (b). The Raman spectroscopy was performed at a low laser power of 5 mW to avoid the formation oxide locally. These curves were fitted with the Gaussian function as shown in figure 1 (b). The first-order phonon frequencies around ~ 230 cm$^{-1}$, 318 cm$^{-1}$, 568 cm$^{-1}$, and 665 cm$^{-1}$ were observed and corresponding to the transverse acoustic (TA), longitudinal acoustic (LA), transverse optical (TO), and longitudinal optical (LO) modes of cubic nanocrystalline TiN respectively.[31] Additionally, the second-order acoustic mode was observed at a frequency of 448 cm$^{-1}$. The first-order acoustic modes arise due to vibrations of Ti atoms near N vacancies, whereas vibrations of N atoms adjacent to Ti vacancies were corresponding to the first-order optical bands. Therefore, the presence of the first-order acoustic and optical modes confirms the presence nitrogen vacancies in the TiN lattice.

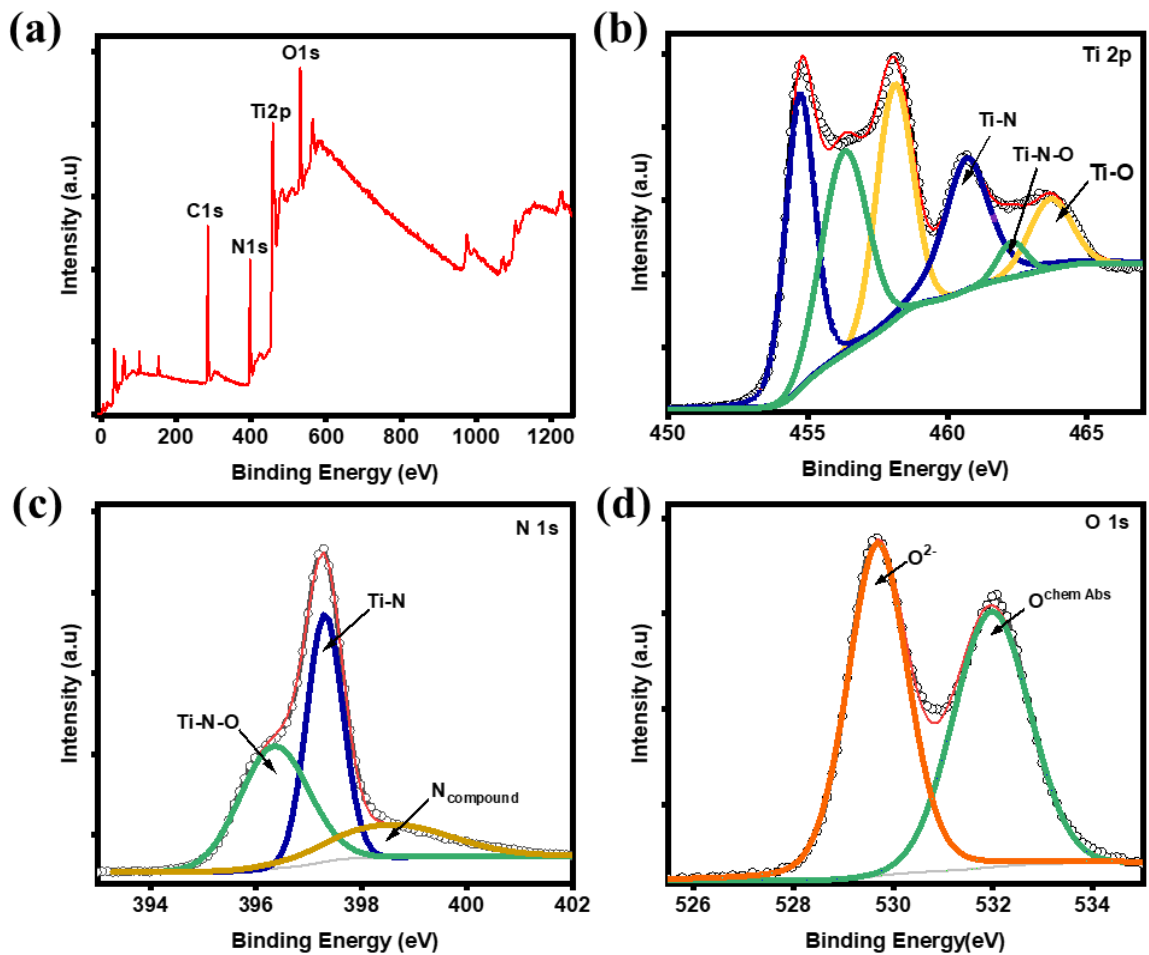

**Figure 2: X-ray photoelectron spectroscopy results of CVA-deposited TiN thin films. (a) the full scan, (b) the Titanium (Ti) 2p orbital binding energy peaks (c,d) Nitrogen (N) and Oxygen (O) 1s orbital binding energy peaks.**

**Figure 2** shows XPS spectra of TiN thin films. The XPS full scan spectra in figure 2 (a) clearly confirms the presence of Ti 2p, N 1s, O1s, and C1s peaks of core energy levels. The fitting and raw XPS spectra are shown in the Figure 2 (b-d). The titanium (Ti 2p) orbital of TiN shows several peaks, corresponding to different Ti bonding. The binding energy peak at 454.78 eV ($p_{3/2}$) and 460 eV ($p_{1/2}$) indicates the presence of Ti-N bonds. Apparently, the oxygen in the TiN film is also seen due to the formation of thin layer of oxynitride on the surface. [7] The formation of the oxide layer gives rise to the peaks at 458.18 eV ($p_{3/2}$) and 463.78 eV ($p_{1/2}$), which indicates the presence of $Ti^{4+}$ and the formation of $TiO_2$. The peak between these two at 456.38 eV($p_{3/2}$) and 462.38 eV($p_{1/2}$) indicates the formation of an intermediate state of $TiO_2$ and TiN i.e, titanium oxynitride ($TiN_xO_y$). In the N(1s) spectrum, peaks were observed at 396.3 eV, 397.2 eV, and 398.5 eV. The highest peak is attributed to the TiN phase, while the other two peaks correspond to $TiN_xO_y$ and nitrogen compounds, respectively. The 1s Oxygen shows two peaks at 529.9 eV and 532.2 eV attributed to $O^{2+}$ and chemically absorbed oxygen ($O^{chem}_{Abs}$) on the surface of TiN.

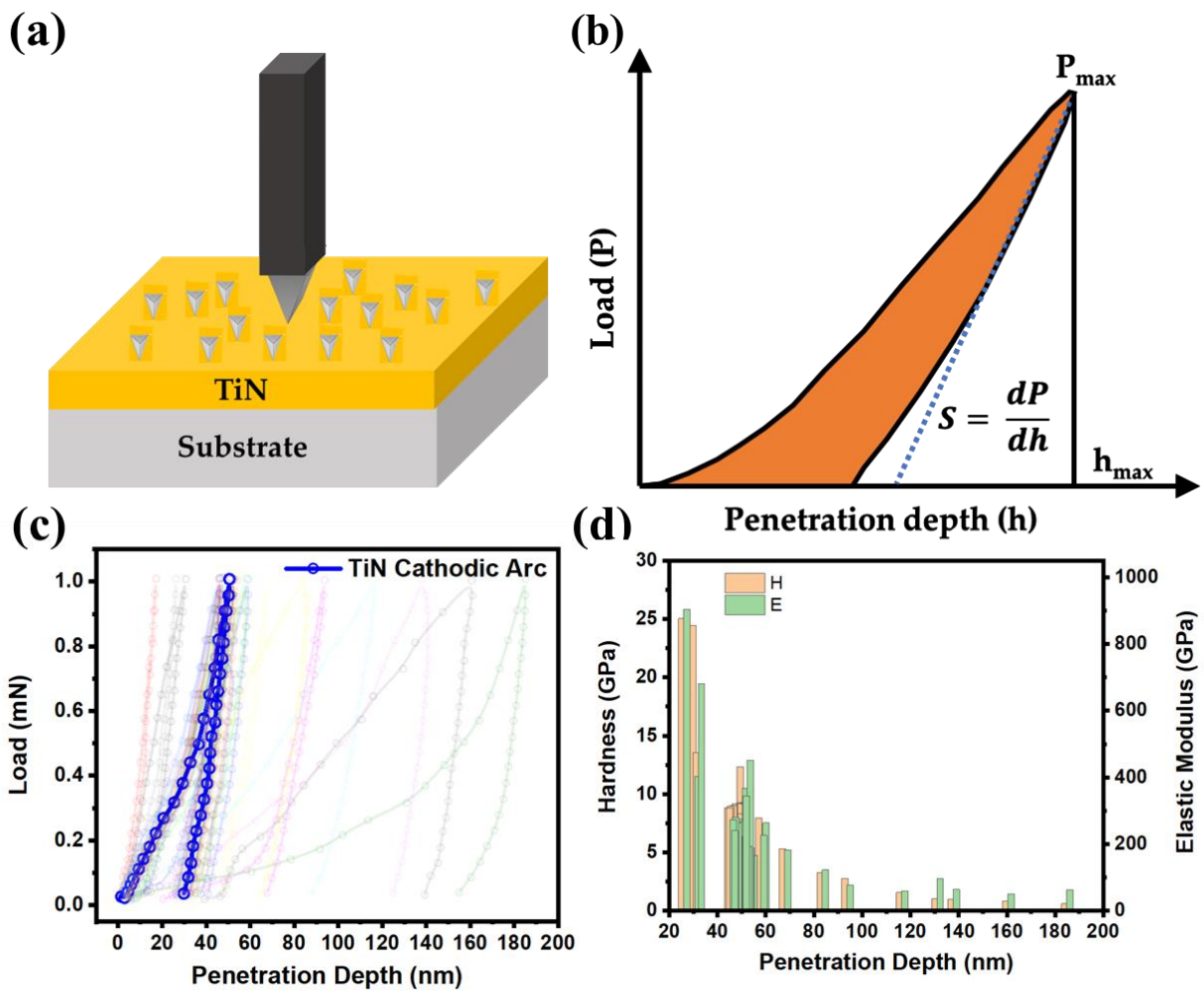

**Figure 3:** (a) Schematic of nanoindentation performed on TiN thin films. (b) Schematic of typical load-displacement curve. (c) The nanoindentation curves obtained at different locations (at few microns spatial distance). (d) the corresponding values of harness (H) on left Y-axis and elastic modulus (E) versus the penetration depth (n).

**Nanoindentation Analysis**

Therefore, we would be interesting to see how these surface defects influences the nanomechanical properties locally. Figure 3 (a-d) shows the nanoindentation measurements to study the nanomechanical properties of CVA-deposited TiN thin films grown on silicon substrates. Figure 3 (a) and 3 (b) illustrate the schematics of the nanoindentation measurement set-up and the typical load versus indent penetration depth obtained to measure the hardness and elastic modulus at nanoscale, which is very similar to the Vickers hardness tester used for measuring microscopic hardness values. In nanoindentation, however, the contact area is estimated by measuring the penetration depth of the indenter into the sample using the diamond indenter tip of known geometry.[32] For the Berkovich indenter tip, the projected contact area $A = 3\sqrt{3}\ h^2\ tan^2\ \theta$, where $h$ is the penetration depth and the angle θ is ~ 65.3° between the tip axis and the faces of the triangular pyramid of the Berkovich indenter, giving $A = 24.5\ h^2$. The hardness and elastic modulus can be calculated [32] using the expressions $H = P/24.5h^2$ and $E = 1/\beta\ \sqrt{\pi}/2\ S/\sqrt{A}$, where β is a constant of ~ 1.034 for the Berkovich indenter and $S$ is the stiffness obtained from the slope of the load-displacement curve as illustrated in figure 3 (b). The nanoindentation curves are obtained at different locations on surface within few microns area and the results are shown in figure 3 (c). It can be observed from these graphs that there is a large spatial variation in the indentation penetration depth (*h*). Apparently, the higher deposition rate of CVA deposited films ~ 1-5 nm/sec can lead to large nitrogen deficiency on the surface. Figure 3 (d) shows the calculated values of the hardness (H) and elastic modulus (E) of both films as a function of the penetration depth. Table 1 summarizes the average values of these parameters.

**Table 1-Nanomechanical properties of TiN films**

|  | **Average $h$ (nm)** | **Average H (GPa)** | **Average E (GPa)** |
|---|---|---|---|
| CVA-TiN | 65 | 8 | 246 |

Since it was not possible to physically see the indent made by nanoindentation method. We have attempted to make indent using the atomic force microscope (AFM) having diamond coated tips. Note these measurements were performed on the films grown at lower nitrogen background pressure compared to the pure TiN films as they are beyond the capability of AFM due to the bending nature of cantilever. To acquire the force-distance (F-D) graphs, as shown in figure S1 (a-b), an AFM tip is brought ~ 1 microns far away from the film surface where there is no interaction with the tip and then pushed against the surface and then finally retracted back to the original position with concurrent recording of the bending of the cantilever (measured as deflection on a detector). Note that the slope in F-D graph does not provide any explicit information about the sample mechanical properties but rather dominated by the spring constant of the cantilever. However, if indentation does occur it will show the hysteresis and a slight change in the slope in the F-D curve. It is interesting to note when the F-D measurement is performed and subsequently the same area is imaged with contact mode AFM, a hysteresis and a clear indent can be seen in the films grown under nitrogen deficient conditions. It was not possible to precisely calculate the hardness value from the penetration depth of AFM tip ~ 150 nm, but our estimated hardness value of ~ 2 GPa agrees nanoindentation measurements, considering the applied load (1V ~ 6300 nN) by AFM divided by the area of the impression (~ 170 $nm^2$) taken from the topographic AFM image [also correlated with SEM image of commercial tip in figure S1 (C)]. Thus, these results confirm that the nanomechanical properties are influenced by the local changes in the nitriding process of these films.

**Finite Element Analysis**

Finite element method has been used for the back analysis of the nanoindentation testing data to obtain the coated material parameters. The real and more complex 3D model was reduced to an axisymmetric 2D model since the model is symmetric with respect to geometry, properties and the loading conditions. This simplifies the analysis procedure, accelerate the back analysis nanoindentation process and reduces the computational cost. The previously reported work also approves that the 2D results matches well with the 3D results for Berkovich indenter[33–35]. The glass substrate of rectangular domain of 400 nm thickness having a 300 nm thick titanium nitride thin film over the surface of glass substrate was modelled, as the geometry shown in Figure 4 (a). In order to decide the suitable width for axisymmetric analysis, representative volume element (RVE) of thin film coated glass surface having a width varying from 250 nm to 1000 nm were generated and a parametric analysis was conducted. The top surface of the thin film was in contact with the indenter having a normal angle of 65.3° and a tip radius of 100 nm. The thin film and the substrate were assumed to be perfectly bonded. A frictionless contact between the indenter and the TiN film was assumed considering the hardness of indenter[36]. The quadrangular shaped second-order elements were used to mesh the geometry. A very fine mesh near the tip of the indenter and around the contact region was created. The FEA solver in ANSYS used full Newton-Raphson method for nonlinear control. The step size was kept very small varying from 1e-06 to 0.1 to obtain the convergence. The large deflection was allowed to accurately model the nonlinear deformation during simulation. The TiN film was modelled as the elastoplastic material with elastic modulus (E) same as predicted from experiments, 246 GPa for CVA deposited film. The plastic deformation of TiN film, was modelled using the Johnson-Cook model in which the flow stress, σ and the equivalent plastic strain, ε are related as:

$$\sigma = (A + B\varepsilon^n)(1 + Cln\dot{\varepsilon}^*)(1 - T^{*m}) \qquad (1)$$

where *A, B, n, C,* and *m* are material constants. A, *B* and n are known as yield stress ($\sigma_y$), hardening modulus and work-hardening exponent, respectively. *C* and *m* are the strain rate hardening and thermal softening coefficient, respectively. For quasistatic conditions with no temperature change, C=0, $T^* = 0$. Hence, the final relation is reduced to:

$$\sigma = B\left[\sigma_y + \left(\frac{\sigma_y}{E}\right)^n\right] \qquad (2)$$

The glass substrate and the diamond indenter were modelled as linear elastic material (see Table 2). A frictionless support was applied at the axis of symmetry and the substrate base was constrained by fixed displacement. The indenter was assigned a displacement boundary condition in a multistep analysis with 2 nm of vertical displacement in each step for loading and unloading conditions. The previously reported work suggest that the yield stress is 1/100[th] of elastic modulus, and the strain hardening exponent for most of metals is considered between 0 to 0.5. A parametric analysis with the assumed values of the hardening modulus and the exponent has been conducted and the simulated nanoindentation curves were compared with the experimental values. The properties used for simulation are mentioned in Table 3.

Figure 4(b) shows the load-indentation curves for different mesh element size varying from 8 nm to 2 nm. This element size was varied for the 100 nm highlighted region near to the tip of the indenter, considering this to be a high stress concentration zone, as shown in Fig. 4(a). As the mesh element size reduces, the number of mesh element increases and hence the computational cost. For 2 nm, mesh element size, it results into a total of 5066 quadrangular second-order elements with 15575 number of nodes. The element quality was also measured using mesh metric 'Element Quality' and it was measured to be a minimum of 0.10179 and an average of 0.557.

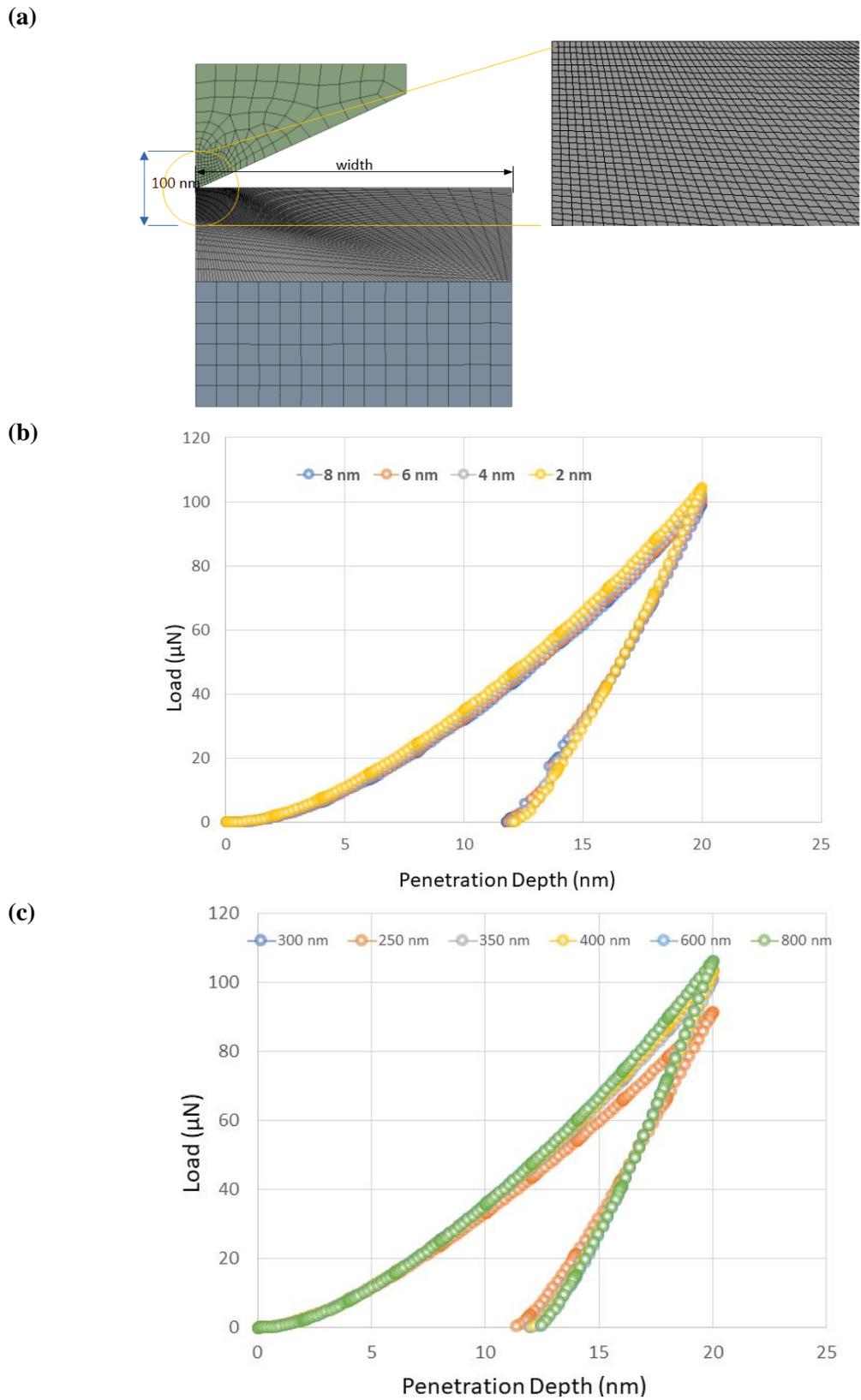

**Figure 4:** (a) Two-dimensional axis symmetric model for nanoindentation. (b) Convergence study for different mesh element size (nm); (c) RVE sensitivity analysis

It can be observed that the load-indentation curves are sensitive to the mesh size and as the element size reduces the results got converged. For 2 nm maximum element size, the simulated load values differ within 2% in the peak load region and therefore this element size is used as the default element length in all simulations. In can also be observed in Figure 4 (c) that the width of RVE has a considerable effect on the load-indentation curves for the 2D axisymmetric modelling. As the RVE width increases more than 400 nm, the predicted load values start converging. Therefore, for all simulation models in this paper, a width of 1000 nm has been considered.

**Table 2**: Mechanical properties of the materials used in the finite element model

|         | Elastic Modulus (GPa) | Poisson's ratio | Yield Stress (MPa) | Hardening exponent | Hardening modulus (MPa) |
|---------|----------------------|-----------------|--------------------|--------------------|--------------------------|
| Glass   | 75                   | 0.23            | -                  | -                  | -                        |
| Diamond | 1141                 | 0.07            | -                  | -                  | -                        |
| TiN     | 246                  | 0.25            |                    |                    |                          |

**Table 3** : Material parameters used for FE simulations for coated TiN film

| Model    | Elastic Modulus (GPa) | Yield Stress ($\sigma_y$) (MPa) | Hardening modulus B (Mpa) | Hardening exponent, n |
|----------|----------------------|-------------------------------|---------------------------|-----------------------|
| model 1  |                      | 1500                          | 2000                      | 0.5                   |
| model 2  |                      | 1500                          | 2000                      | 0.1                   |
| model 3  |                      | 1500                          | 1500                      | 0.5                   |
| model 4  | 246                  | 1000                          | 1500                      | 0.5                   |
| model 5  |                      | 2500                          | 1500                      | 0.1                   |
| model 6  |                      | 2500                          | 6000                      | 0.1                   |
| model 7  |                      | 3500                          | 10000                     | 0.1                   |
| model 8  |                      | 3500                          | 15000                     | 0.1                   |
| model 9  | 400                  | 3500                          | 10000                     | 0.1                   |
| model 10 | 200                  | 1500                          | 2000                      | 0.5                   |
| model 11 | 200                  | 1000                          | 1500                      | 0.5                   |
| model 12 |                      | 600                           | 1000                      | 0.5                   |
| model 13 |                      | 4500                          | 10000                     | 0.1                   |
| model 14 | 246                  | 2500                          | 25000                     | 0.1                   |
| model 15 |                      | 2500                          | 30000                     | 0.1                   |
| model 16 |                      | 2500                          | 25000                     | 0.2                   |

Figure 5 shows the simulated load-indentation curves for 16 different models. The simulated curves are compared with the experimental curves obtained in case of CVA deposited coating. It can be observed that the indentation depth increases as the yield stress and hardening modulus reduces. In case of model 12, the yield stress and hardening modulus are 600 MPa and 1000 MPa respectively, that results into a 146 nm indentation at a peak load of 0.972 mN. A lower value of yield stress initiates plastic deformation earlier. The strain hardening behaviour of the coated material depends on the hardening modulus and the hardening exponent. Therefore, the lower values of hardening modulus results into increased plastic strain. Another observation that can be made from the simulated plots is that the lower values of hardening exponent results into high values of flow stresses and therefore more load and less indentation depth. The experimental plots for Exp-8 to 10, are best represented by FEA model-11 and model-12. The high penetration depth in case of experiments 8-10 shows that the coating in that region is more ductile with low yield stress in between 600 to 1000 MPa, hardening modulus in between 1000-1500 MPa and hardening exponent 0.1. This, further supports our previous conclusions from experiments that the region of CVA deposited film is grown under nitrogen deficient conditions leading to more ductile behaviour. A clustering of experimental curves can be observed in the region with penetrating depth less than 40 nm. This low penetration depth curves are best represented by models having yield stress around 2500 MPa and a high hardening modulus close to 25000 MPa.

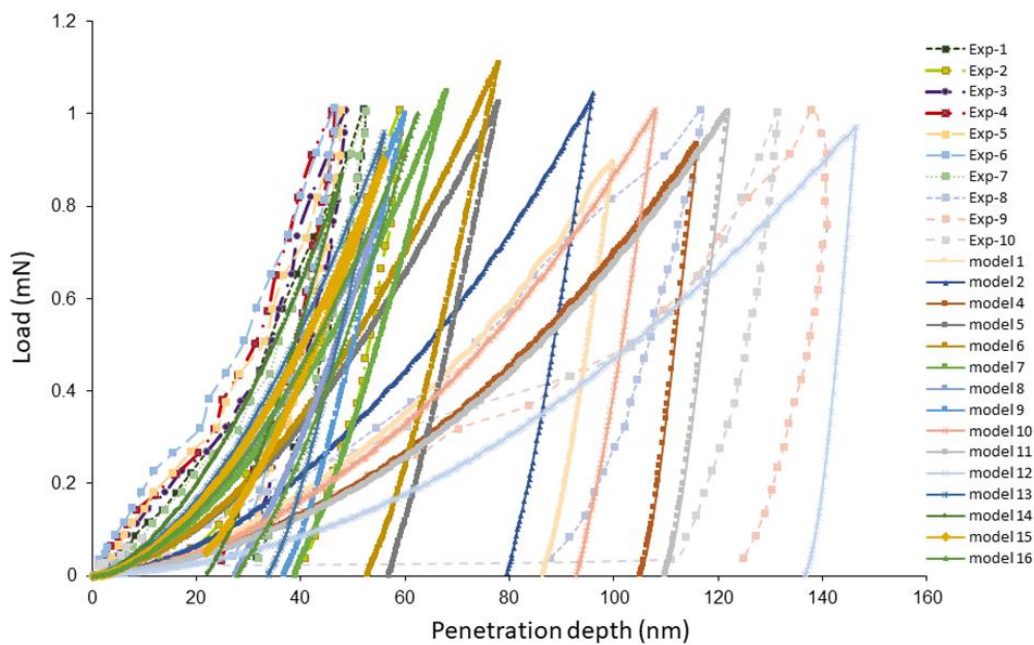

**Figure 5: Nanoindentation curves resulted from FEM analysis for CVA deposited coating with 16 different models with 600 MPa ≤σ$_y$ ≤ 3500 MPa and 0.1≤n≤0.5**

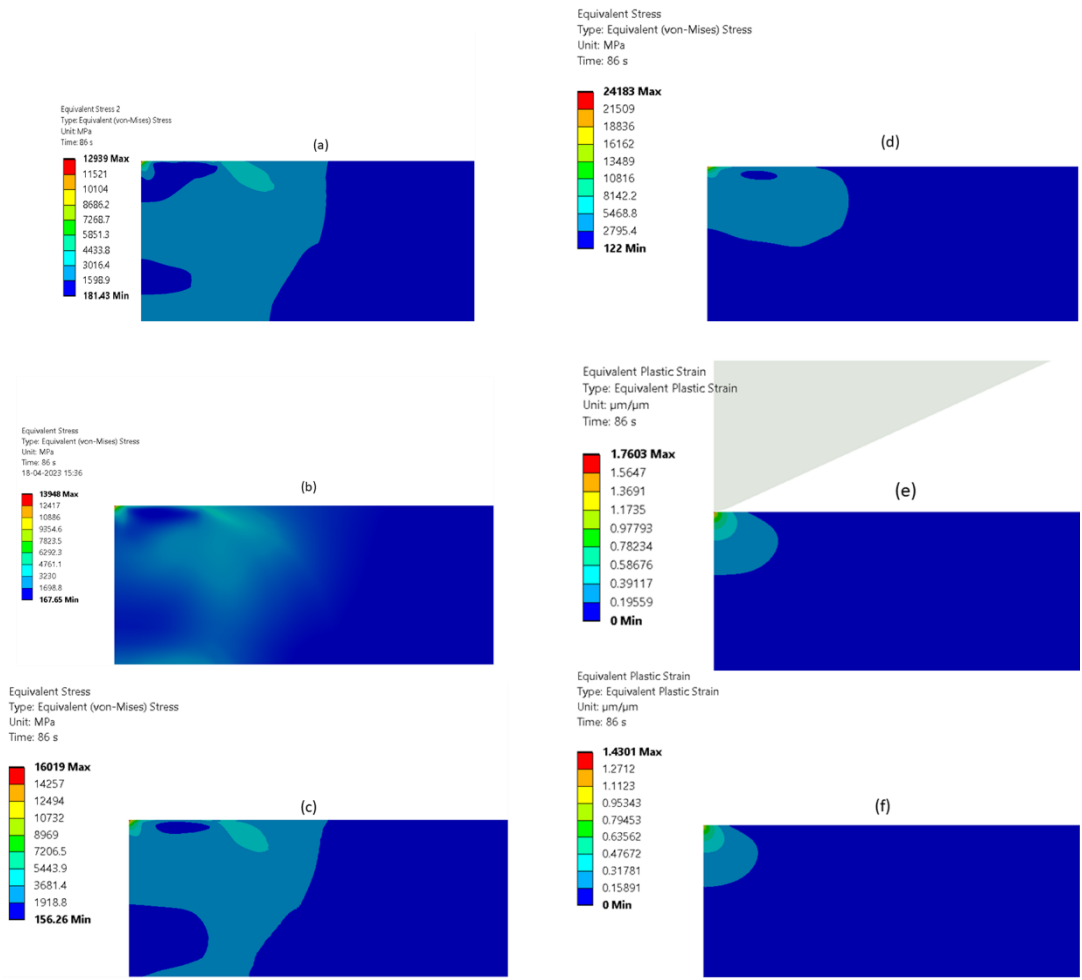

**Figure 6: von- Mises stress (MPa) contours for TiN thin film (a) model 3, (b) model 5, (c) model 6, (d) model 7, (e) Equivalent Plastic strain model 6, (f) equivalent plastic strain model 7**

Figure 6 (a-d) shows the von-Mises stress for four different models: model 3, model 5, model 6 and model 7. The stress concentration increases as the yield stress increases. Model 3 has a yield stress of 1500 MPa and hence the plastic deformation starts earlier comparing to the model 7 which has a yield stress of 3500 MPa. A higher value of equivalent plastic strain for model 6 [Figure 6 (e)] comparing to model 7 [Figure 6 (f)] is due to a lower value of yield stress for model 6. In order to understand the effects of substrate, the nanoindentation simulations of TiN thin film coated on glass and silicon substrate are compared in Figure 7. It can be seen from the nanoindentation curves that the elastic modulus of both the curves are same, however the hardening coefficient of silicon substrate deposited TiN film is slightly higher than that of glass substrate deposited TiN film. The strain hardening behaviour post yielding depends on the dislocations movement and grain boundaries formation, which is affected by the interface of thin film and substrate and a soft substrate like silicon results high hardening modulus.

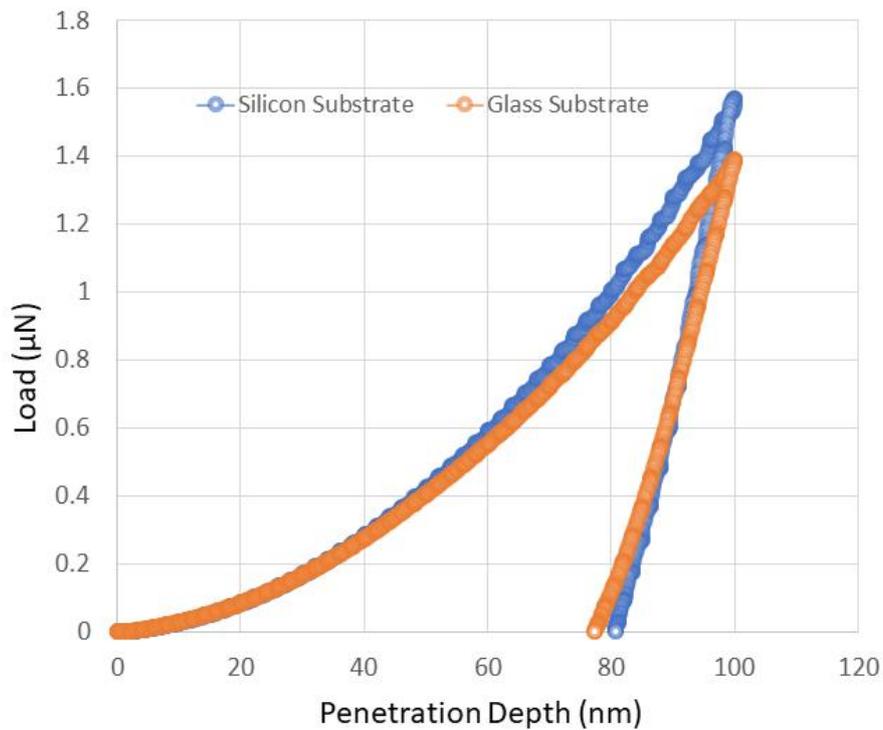

**Figure 7: Nanoindentation curve for glass and silicon substrate for model 4**

**Conclusion**

In conclusion, the nanoscale spatial inhomogeneity in mechanical properties of TiN thin films were investigated using nanoindentation and atomic force microscopy and spectroscopy. The local nanoindentation load versus penetration depth graphs were simulated using finite element method (FEM), that confirms the films having local nitrogen deficiency are ductile with low yield stress and hardening modulus. This investigation further opens the new opportunities of modelling the nanoscale system using FEM analysis and predict the nanoindentation results.

**Acknowledgements**

ASR would like to acknowledge SERB Core Research Grant CRG/2021/001l36 for the financial support. ASR would like to thank Prof. B. S. Satyanarayana for the discussions.

**Author contributions**

ASR and NKS conceived the idea and executed overall experimental design, results analysis and wrote the original draft of manuscript. NKS performed FEA modelling and analysis. AR grew samples, performed XRD and Raman spectroscopy and analysed the results. OSP helped in the nanoindentation measurements and the interpretation of these results. ASR supervised the overall work and reviewed the final manuscript.

**Competing interests**

The authors declare no competing interests.

**Data availability**

All data generated or analysed during this study are included in this published article and its supplementary information files.